\documentclass{PoS}

\title{Overlap/Domain-wall reweighting}

\ShortTitle{Overlap/Domain-wall reweighting}

\author{JLQCD Collaboration: 
        \speaker{H.~Fukaya}$^a$\thanks{E-mail: hfukaya@het.phys.sci.osaka-u.ac.jp},
        S.~Aoki$^{b}$,
        G.~Cossu$^{c}$,
        S.~Hashimoto$^{c,d}$,
        T.~Kaneko$^{c,d}$,
        J.~Noaki$^{c}$
        \\
        \\
        \\
        \llap{$^a$}
        Department of Physics, Osaka University, 
        Toyonaka, Osaka 560-0043, Japan
        \\
        \llap{$^b$}
        Yukawa Institute for Theoretical Physics, Kyoto University, Kyoto 606-8502, Japan
        \\
        \llap{$^c$}
        High Energy Accelerator Research Organization (KEK),
        Tsukuba 305-0801, Japan 
        \\
        \llap{$^d$}
        School of High Energy Accelerator Science,
        The Graduate University for Advanced Studies (Sokendai),
        Tsukuba 305-0801, Japan
}

\abstract{We investigate the eigenvalues of nearly chiral lattice Dirac operators constructed with five-dimensional implementations. Allowing small violation of the Ginsparg-Wilson relation, the HMC simulation is made much faster while the eigenvalues are not significantly affected. We discuss the possibility of reweighting the gauge configurations generated with domain-wall fermions to those of exactly chiral lattice fermions.}

\FullConference{31st International Symposium on Lattice Field Theory - LATTICE 2013\\
		July 29 - August 3, 2013\\
		Mainz, Germany}

\begin{document}

\section{Motivation}

The JLQCD collaboration started a new project of simulating lattice
QCD with nearly chiral lattice fermions at fine lattice spacings,
utilizing the machines at KEK, IBM Blue Gene/Q and Hitachi SR16000,
installed in 2012.
Our goal is to realize precise calculations for quark flavor physics,
which are necessary to match the planned precise experiments, such as
SuperKEKB/Belle II.
We plan to generate lattices with the cutoff $1/a$ between 2.4 and
4.8~GeV, aiming at reducing discretization effects for heavy quark as
much as possible.
As the required precision reaches the percent level, we expect that
the controlled chiral symmetry would become more important.
It doesn't necessarily mean that we require {\it exact} chiral
symmetry provided by the overlap fermion
\cite{Neuberger:1997fp}, but rather we would
accelerate the simulation by allowing some small violation with the
domain-wall-type formulation
\cite{Kaplan:1992bt,Shamir:1993zy,hep-lat/0005002, hep-lat/0209153, Brower:2012vk}.

Theoretically, the overlap and domain-wall formulations are similar,
{\it i.e.} just different expressions of the chirally symmetric Dirac
operator that satisfies the Ginsparg-Wilson (GW) relation. 
When we perform
numerical simulations, their difference is even less rigid, since
neither expression can exactly satisfy the GW relation.
For the following discussion, let us {\it define} the overlap fermion
as the one that satisfies the GW relation at $10^{-8}$ level or less,
while the domain-wall fermions are those having the violation as large
as $10^{-3}$, which is roughly 1--5~MeV level and could be significant
near the physical point.

There are other issues to be considered. The computational cost is one
of those. The domain-wall fermion is significantly faster than the
overlap fermion when doing the Hybrid Monte Carlo (HMC) simulation.
Topological tunneling is easier with domain-wall. But the eigenmode
calculation is easier with overlap, because one can use chiral
projection to reduce the functional space.

In this work, we address a possibility of using both formulations,
{\it i.e.} performing the configuration generation with domain-wall
and reweight the resulting configurations according to the overlap
fermion determinant. If this is feasible, we can utilize the advantages
of both formulations. To be specific, our configuration generation is
already on-going with a variant of the domain-wall fermions, {\i.e.}
scaled Shamir kernel on stout-smeared gauge links. The question is
thus to find a practically optimal overlap implementation among many choices of
the kernel and the sign function approximation.

\section{Overlap vs. Domain-wall}

The domain-wall fermion action 
was originally proposed by Kaplan \cite{Kaplan:1992bt}
and many variations have so far been proposed 
\cite{Shamir:1993zy,hep-lat/0005002, hep-lat/0209153}. 
Brower {\it et al.} \cite{Brower:2012vk}
summarized them in a compact form,
\begin{equation}
  \label{eq:gdw_action}
  S_{GDW} = \sum_{x,s} \bar{\psi} D_{GDW}^5 \psi,
\end{equation}
with a 5-dimensional operator
\begin{equation}
  \label{eq:gdw}
  D_{GDW}^5=\left(
    \begin{array}{cccccc}
      \tilde{D}^1 & -P_- & 0 & ... & 0 & m P_+\\
      -P_+ & \tilde{D}^2 & -P_- & 0 & ... & 0 \\
      0 & -P_+ & \tilde{D}^3 & \ddots & 0 & \vdots\\
      \vdots & 0 & \ddots & \ddots & \ddots & \vdots\\
      0 & ... & \ddots & -P_+ & \tilde{D}^{L_s-1} & -P_-\\
      mP_- & 0 & ... & 0 & -P_+ & \tilde{D}^{L_s}
    \end{array}
  \right),
\end{equation}
where the rows and columns denote the 5th direction (labeled by $s=1,2,\cdots L_s$),  and
\begin{eqnarray}
  \tilde{D}^s & = & (D_-^s)^{-1} D_+^s, \\
  D_+^s & = & 1 + b_s D_W(-M), \\
  D_-^s & = & 1 - c_s D_W(-M),
\end{eqnarray}
with some ($s$-dependent) constants $b_s$ and $c_s$.
Here, $m$ is the fermion mass that we should have 
in the 4-dimensional effective Lagrangian, $P_{\pm}$ denotes
the chiral projection operators, and $D_W(-M)$ 
is the 4-dimensional Wilson Dirac operator with a negative mass $-M$.

This form indicates that there are many options in defining the domain-wall fermions,
by choosing the 5th coordinate dependent parameters.
For example, the standard domain-wall fermion is obtained
by setting $b_s=1$, and $c_s=0$ for all $s$.
When we consider the 4-dimensional effective 
operator (with the Pauli-Villars field), however,
one can see that their difference is not essential:
it is only a different way to approximate the same operator.

To see this, we make use of an equality 
of $\det D_{GDW}^5(m)/D_{GDW}^5(m=1) $ (after some matrix manipulations)
to the determinant of the 4-dimensional operator 
\begin{eqnarray}
D^{4}_{GDW}&=&
    \frac{1+m}{2}+\frac{1-m}{2}\gamma_5
    \frac{
      T_1^{-1}T_2^{-1}\cdots T_{L_s}^{-1}-1}{
      T_1^{-1}T_2^{-1}\cdots T_{L_s}^{-1}+1},
\end{eqnarray}
where the ``transfer matrix'' $T_s$ is given by
\begin{eqnarray}
  T_s^{-1} 
  & = &
  \left[
    1-\gamma_5\frac{(b_s+c_s)D_W}{2+(b_s-c_s)D_W}
  \right]^{-1}
  \left[
    1+\gamma_5\frac{(b_s+c_s)D_W}{2+(b_s-c_s)D_W}
  \right].
\end{eqnarray}
If we set the parameters as
\begin{equation}
  \begin{array}{cc}
    b_s+c_s=b\omega_s, & b_s-c_s=c
  \end{array}
\end{equation}
with new parameters $b$, $c$ and $\{\omega_s\}$, we may define the
M\"obius kernel
\begin{equation}
  \label{eq:HM}
  H_M = \gamma_5 \frac{b D_W}{2+cD_W}, 
\end{equation}
which interpolates between the Borici (or Wilson) kernel $H_W$ ($b=2$,
$c=0$) and the Shamir (or DW) kernel $H_T$ ($b=c=1$),
and the transfer matrix becomes
\begin{equation}
  T_s^{-1} = \frac{1+\omega_s H_M}{1-\omega_s H_M}.
\end{equation}
Thus, the different expression in the 5-dimensional $D_{GDW}^5(m)$
results in the different choices of the approximation for the
sign function,
\begin{equation}
  \mathrm{sgn}_{\mathrm{rat}} = 
  \frac{1-\prod_s T_s}{1+\prod_s T_s}.
\end{equation}
It is known that the optimal choice is that of Zolotarev, 
the optimal rational approximation in a given interval of the eigenvalue of $H_M$.
The detailed expression is given in Ref.~\cite{hep-lat/0209153}.

Numerically, the overlap Dirac operator is a refinement of 
this 4-dimensional expression of the domain-wall operator.
The simplest way to achieve this is to use a large $L_s$.
But, since the sign function becomes worse near 
the zero-eigenvalue of the kernel, 
it is well-known that the exact treatment of the kernel low-modes, 
\begin{eqnarray}
D_{ov} &=& \frac{1}{2}\underbrace{\sum_{|\lambda_i|< \lambda_{th}} \left(1+ \gamma_5 {\rm sgn}\lambda_i\right)
|\lambda_i\rangle\langle \lambda_i |}_{\mbox{Low modes}} 
+ D_{GDW^\prime}^{4}\underbrace{\left(1-
\sum_{|\lambda_i|< \lambda_{th}}|\lambda_i\rangle\langle \lambda_i |\right)}_{\mbox{High modes}},
\end{eqnarray}
is useful to reduce the chiral symmetry violation.
Here, $\lambda_i$'s denote the eigenvalues of the hermitian 
kernel operator $H_M$, whose absolute values are below 
some threshold $\lambda_{\rm th}$.
Then the overlap fermion action can be implemented
with a reasonable depth in the 5-th direction, $L_s\sim {\cal O}(10)$.
Note here that we have added a ``prime'' to the 
domain-wall operator, $D_{GDW^\prime}^{4}$, 
since different implementation of the domain-wall operator
can be chosen for the valence overlap quarks.

Let us here choose $D_{GDW^\prime}^{4}=D_{GDW}^{4}$ and
multiply its inverse to $D_{ov}$,
\begin{eqnarray}
(D_{GDW}^{4})^{-1}D_{ov} &=& 1 + \sum_{|\lambda_i|< \lambda_{th}}
((D_{GDW}^{4})^{-1}D_{ov}-1)|\lambda_i\rangle\langle \lambda_i |.
\end{eqnarray}
Then, one can see why this reweighting should work.
The reweighting factor is almost unity, and only (hopefully) 
small correction comes from the low-mode subvolume of the kernel $H_M$.

\section{Preliminary lattice results}

In this section, we report on our preliminary lattice studies
on a small lattice. 
The numerical simulations are performed using
the multipurpose C++ code IroIro++\cite{Cossu}.

For the domain-wall fermion action as sea quarks,
we have performed various test runs.
The details are shown in Refs.~\cite{Kaneko,Noaki,Hashimoto}.
After these tests, our choice, 
what we believe optimal for 
the HMC updates, is the one with 
the scaled Shamir kernel 
$H_M= \gamma_5 \frac{2 D_W}{2+D_W}=2H_T$,
and the polar approximation of the sign function: 
${\rm sgn}(H_M)=\frac{(1+H_M)^{L_s}-(1-H_M)^{L_s}}{(1+H_M)^{L_s}+(1-H_M)^{L_s}}$.
With $L_s=12$, and other parameter choices (below), 
we can achieve $\sim 10^{-4}$ precision of the chiral symmetry,
or the residual mass $m_{\rm res}\sim 0.5$ MeV.
 
In this work, we concentrate on our test runs on 
a $16^3\times 32$ lattice.
For the gauge action, we employ the Symanzik improved 
gauge action with $\beta=4.17$.
For the quark part of the action, we use three steps
of the stout smearing.
With the strange quark mass $m_s=0.039$, which is almost at the physical point,
the up and down quark mass is set at $m_{ud}\sim 0.7 m_s$.
Our estimate for the lattice spacing is around 2.4 GeV
(which is determined from the scale ``$t_0$'' of the Wilson Flow);
the physical lattice size is $\sim 1.3$ fm.


For the valence overlap quarks, 
we try two different approximations of the sign functions, 
Zolotarev and polar approximations,
different values of the threshold $\lambda_{th}$
under which the low-modes of the kernel are exactly treated,
and different values of $L_s$.

\begin{figure}[bt]
  \centering
  \includegraphics[width=9.5cm]{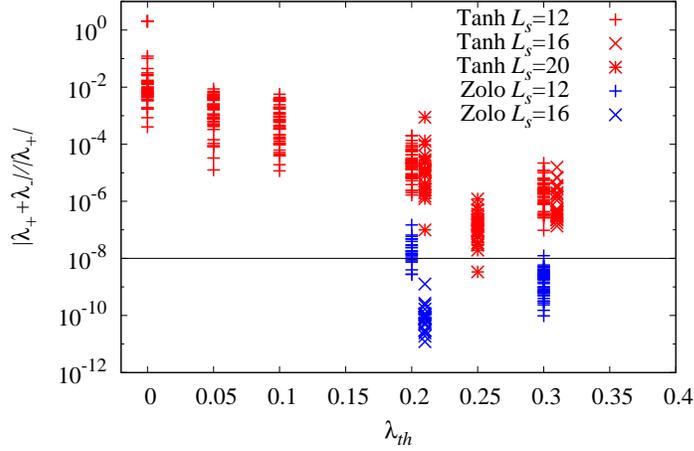}
  \caption{Check of the chirality by the degeneracy of eigenpairs ($\lambda_+, \lambda_-$).}
  \label{fig:chirality}
\end{figure}

To examine the chiral symmetry violation, 
we use the eigenvalues of the hermitian operator $H_{ov}=\gamma_5 D_{ov}$. 
If the Ginsparg-Wilson relation 
is exactly satisfied, any positive eigenvalue of $H_{ov}$ (denoted by $\lambda_+$) makes
a pair with another negative eigenvalue, $\lambda_-$ that has the same magnitude.
Therefore, the ratio
\begin{eqnarray}
\left|\frac{\lambda_+ + \lambda_-}{\lambda_+}\right|,
\end{eqnarray}
is a good indicator of the chirality.
Figure~\ref{fig:chirality} shows a comparison 
among different implementation of the overlap Dirac operator.
The use of the Zolotarev approximation in the range $[0.3,1.65]$,
with the low-mode threshold $\lambda_{\rm th}=0.3$,
gives a ${\cal O}(10^{-8})$ level of precision for the chirality
with $L_s=12$.
In the following, we use this set-up as the implementation
of the valence overlap Dirac operator.

To calculate the reweighting factor,
\begin{eqnarray}
R &=& \left(\frac{\det \gamma_5 D_{ov}(m_{ud})}{\det \gamma_5 D_{GDW}^4(m_{ud})}\right)^2,
\end{eqnarray}
we use the stochastic estimator with $60$--$300$ Gaussian noises.
Note here that the effect of the strange quark determinant is
neglected (quenched) in this preliminary study, 
although it is included in the configuration generations.
We also calculate 100 lowest eigenvalues of $\gamma_5 D_{ov}(m_{ud})$,
and $\gamma_5 D_{GDW}^4(m_{ud})$, to see the effect of the low-mode part ($R_{\rm low}$).

\begin{figure}[tb]
  \centering
  \includegraphics[width=9.5cm]{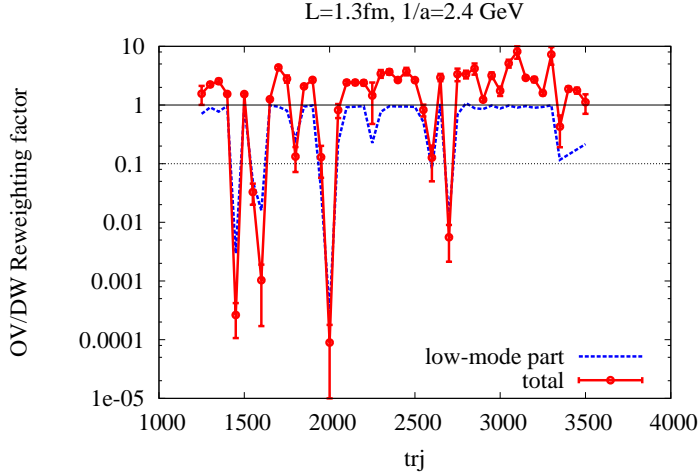}
  \caption{The HMC history of the reweighting factor.}
  \label{fig:reweight}
\end{figure}

Figure~\ref{fig:reweight} shows the HMC history of the 
reweighting factor. Here, red filled circles 
denote the total contribution, while blue dashed lines
show their low-mode part.
From this figure, we find that
the low-mode and high-mode contributions have different tendencies.
For 80\% of configurations, the low-mode part $R_{\rm low} \lesssim 1$ 
but occasionally the reweighting factor becomes exceptionally small.
On the other hand, the high mode contribution 
always makes $R$ bigger, except for the cases 
with a very small low-mode part.
We may conclude that the high-mode or UV fluctuation
makes $R$ larger, while the low-mode or IR fluctuation
makes $R$ smaller.

\begin{figure*}[tb]
  \centering
  \includegraphics[width=6cm]{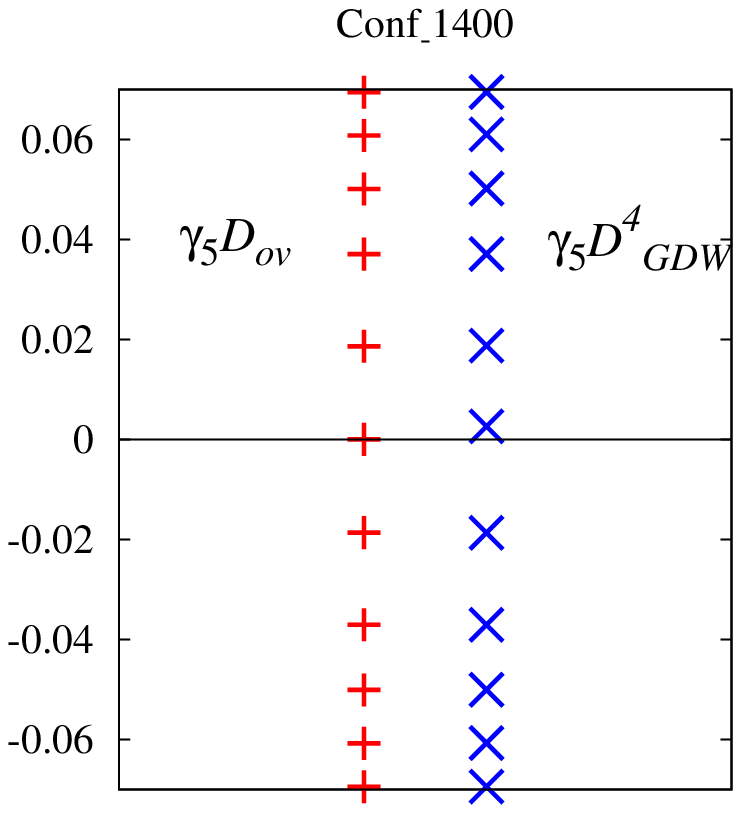}
  \includegraphics[width=6cm]{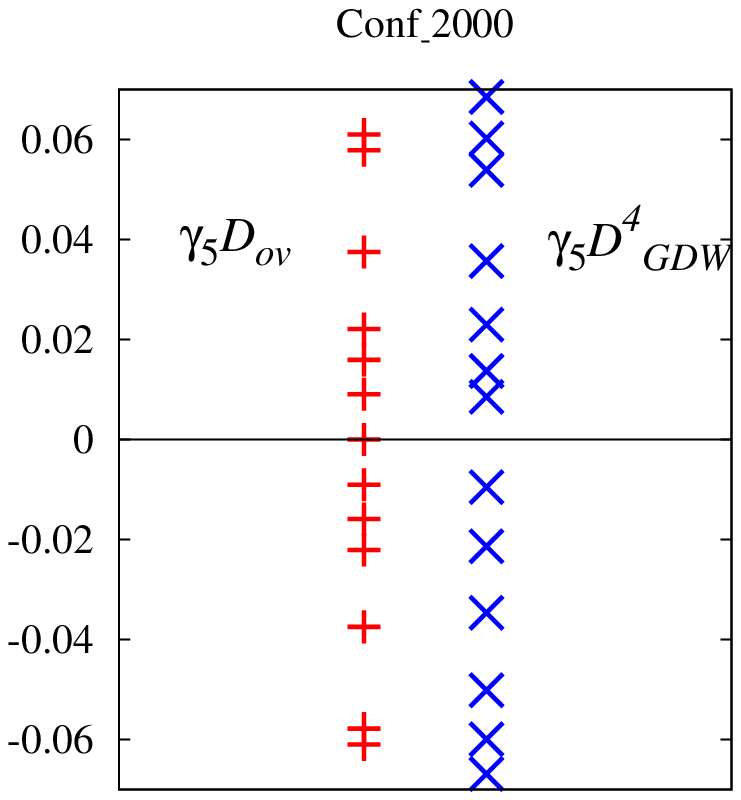}
  \caption{Comparison of the low-mode spectrum 
of $\gamma_5D_{ov}$ (red) and that of $\gamma_5D_{GDW}^4$ (blue).
The left panel shows the case with $R=1.54(6)$, for which 
the low-mode spectrum is similar to each other.
The right panel, the one with $R=0.00009(8)$,
shows a big mismatch, the number of the zero-modes looks different.}
\label{fig:Qmismatch}
\end{figure*}

For the cases with the exceptionally small reweighting factor,
we observe a big mismatch in the low-mode spectrum:
their topological charge looks different, as shown in 
Fig.~\ref{fig:Qmismatch}.
For configurations having $R\sim 1$, as the left panel showing the case with $R=1.54(6)$,
we observe a good agreement in the low-mode spectrum.
However, for those having very small $R$,
as the right panel with $R=0.00009(8)$ shows,
they look quite different.
In particular, the zero-mode of $\gamma_5D_{ov}$ is missing.

We also notice that the high-mode fluctuation is
larger than our naive estimate $R\sim 1+{\cal O}(m_{\rm res})$.
This is not due to the difference between 
$D_{GDW^\prime}^{4}$ and $D_{GDW}^{4}$: 
replacing $D_{GDW^\prime}^{4}$(Zolotarev) with $D_{GDW}^{4}$ (poler)
changes each reweighting factor only by 1\%.
We note that similar results 
have been reported by other groups~\cite{Ishikawa:2010tq, Ishikawa:2013rxa}.

\section{Summary and discussion}

Theoretically the overlap and (generalized) domain-wall fermions 
are merely two different ways of approximations of the same action.
In particular, when we use the same numerical implementation
of the sign function for the high-mode part of 
the kernel $H_M$, the difference only comes from
a limited subspace, the low-mode part of $H_M$.
Since the domain-wall fermion action makes the HMC updates 
numerically cheaper, while the overlap fermion 
allows us a cleaner analysis in the measurement,
we have attempted a reweighting of their determinant, 
expecting a small fluctuation in the reweighting factors.

On a small pilot lattice ($L\sim 1.3$ fm) generated 
using the domain-wall fermion action (with the scaled Shamir kernel)
we have computed the overlap/domain-wall 
reweighting factor $R$.
For 70\% of our configurations, $R$ is close to unity. 
However, the remaining 30\% show a
quite small $R<0.1$ or a large $R>4$.
Separating 100 lowest mode's contribution
from the others, we found that 
the low-mode contribution is responsible for
the small reweighting factor, especially, when there is
a mismatch in the topological charge,
while the high-mode contribution is
responsible for the large reweighting factor.
We may need to suppress both fluctuations
for a successful reweighting on larger lattices.\\

We thank N.~Christ, S.~D\"urr, and S.~Schaefer for useful discussions.
Numerical simulations are performed on the IBM System Blue Gene
Solution at High Energy Accelerator Research Organization
(KEK) under a support of its Large Scale Simulation Program (No.12/13--04).
This work is supported in part by the Grant-in-Aid of the
  Japanese Ministry of Education
  (No. 21674002, 25287046, 25800147), 
and SPIRE (Strategic Program for Innovative Research).


\end{document}